\documentstyle[12pt,epsf]{article}

\newcommand{\newc}{\newcommand}

\newc{\vs}{{\it vs.}}
\newc{\msusy}{M_{\rm SUSY}}
\newc{\wgratio}{x}
\newc{\ssqthw}{\sin^2\theta_W}
\newc{\Rb}{R_b}
\newc{\mone}{M_1}
\newc{\mtwo}{M_2}
\newc{\alphasmz}{\alpha_s(\mz)}
\newc{\alphasmzmin}{\alphas^{\rm min}(\mz)}

\newc{\slepton}{{\tilde{l}}}    \newc{\mslepton}{m_\slepton}
\newc{\supq}{{\tilde{u}}}
\newc{\sdown}{{\tilde{d}}}
\newc{\selectron}{{\tilde{e}}}
\newc{\sneutrino}{{\tilde{\nu}}}
\newc{\squark}{{\tilde{q}}}     \newc{\msquark}{m_\squark}
\newc{\higgsino}{{\tilde{H}}}

\newc{\mtop}{\mt}
\newc{\btau}{$b$--$\tau$}
\newc{\bino}{\widetilde B}	\newc{\mbino}{m_{\bino}}
\newc{\wino}{\widetilde W}	\newc{\mwino}{m_{\wino}}
\newc{\beq}{\begin{equation}}
\newc{\eeq}{\end{equation}}
\newc{\bea}{\begin{eqnarray}}
\newc{\eea}{\end{eqnarray}}
\newc{\onehalf}{\frac{1}{2}}
\newc{\gsim}{\lower.7ex\hbox{$\;\stackrel{\textstyle>}{\sim}\;$}}
\newc{\lsim}{\lower.7ex\hbox{$\;\stackrel{\textstyle<}{\sim}\;$}}
\newc{\alphas}{\alpha_s}
\newc{\tanb}{\tan\beta}
\newc{\mz}{m_Z}         \newc{\mw}{m_W}
\newc{\mhalf}{m_{1/2}}
\newc{\mzero}{m_0}
\newc{\muzero}{\mu_0}
\newc{\sgnmu}{{\rm sgn}\,\muzero}
\newc{\azero}{A_0}
\newc{\Atop}{A_t}
\newc{\bzero}{B_0}
\newc{\mt}{m_t}
\newc{\mb}{m_b}
\newc{\mtau}{m_\tau}
\newc{\mq}{m_q}
\newc{\htop}{h_t}
\newc{\hbot}{h_b}
\newc{\htau}{h_\tau}
\newc{\mtpole}{M_t}
\newc{\mbpole}{M_b}
\newc{\mqpole}{M_q}
\newc{\mgut}{M_X}
\newc{\mx}{\mgut}
\newc{\alphax}{\alpha_X}
\newc{\ie}{{\it i.e.}}
\newc{\etal}{{\it et al.}}
\newc{\eg}{{\it e.g.}}
\newc{\etc}{{\it etc.}}
\newc{\hh}{{h^0}}
\newc{\mhh}{m_\hh}
\newc{\hH}{{H^0}}
\newc{\mhH}{m_\hH}
\newc{\hA}{{A^0}}
\newc{\mhA}{m_\hA}
\newc{\hpm}{{H^\pm}}
\newc{\mhpm}{m_\hpm}
\newc{\stp}{{\widetilde t}}
\newc{\stopl}{{\stp_L}}         \newc{\mstopl}{m_{\stopl}}
\newc{\stopr}{{\stp_R}}         \newc{\mstopr}{m_{\stopr}}
\newc{\stau}{{\widetilde\tau}}
\newc{\gluino}{{\widetilde g}}  \newc{\mgluino}{m_{\gluino}}
\newc{\MS}{{\rm\overline{MS}}}
\newc{\DR}{{\rm\overline{DR}}}
\newc{\ev}{{\rm\,eV}}
\newc{\mev}{{\rm\,MeV}}
\newc{\gev}{{\rm\,GeV}}
\newc{\tev}{{\rm\,TeV}}
\newc{\bsg}{BR$(b\to s\gamma)$}
\newc{\abundchi}{\Omega_\chi h_0^2}
\newc{\mchi}{m_\chi}
\newc{\mcharone}{m_{\charone}}
        \newc{\charone}{\chi_1^\pm}
\newc{\mneutone}{m_{\neutone}}  \newc{\neutone}{\chi^0_1}
\newc{\mneuttwo}{m_{\neuttwo}}  \newc{\neuttwo}{\chi^0_2}

\def\PLB#1#2#3{{\rm Phys. Lett.} {\bf B#1} (19#2) #3}

\def\PRD#1#2#3{{\rm Phys. Rev.} {\bf D#1} (19#2) #3}

\def\IJMPA#1#2#3{Int. J. Mod. Phys. {\bf A#1} (19#2) #3}

\begin{document}

\setlength{\baselineskip}{3.0ex}

\newcommand{\st}{\scriptstyle}
\newcommand{\sst}{\scriptscriptstyle}
\newcommand{\mco}{\multicolumn}
\newcommand{\epp}{\epsilon^{\prime}}
\newcommand{\vep}{\varepsilon}
\newcommand{\ra}{\rightarrow}
\newcommand{\ppg}{\pi^+\pi^-\gamma}
\newcommand{\vp}{{\bf p}}
\newcommand{\ko}{K^0}
\newcommand{\kb}{\bar{K^0}}
\newcommand{\al}{\alpha}
\newcommand{\ab}{\bar{\alpha}}
\def\be{\begin{equation}}
\def\ee{\end{equation}}
\def\bea{\begin{eqnarray}}
\def\eea{\end{eqnarray}}
\def\CPbar{\hbox{{\rm CP}\hskip-1.80em{/}}}
\begin{titlepage}
\renewcommand{\thefootnote}{\fnsymbol{footnote}}

\begin{center} \Large
{\bf Theoretical Physics Institute}\\
{\bf University of Minnesota}
\end{center}

\begin{flushright}
TPI-MINN-95/26-T\\
UMN-TH-1408-95\\
hep-ph/9509273\\
September 1995
\end{flushright}
\vspace{1cm}

\centerline{\large\bf GAUGE COUPLING UNIFICATION
VS. SMALL $\alpha_s\approx0.11$}

\vspace*{6.0ex}
\begin{center}{\large LESZEK ROSZKOWSKI\footnote{
Talk presented
at SUSY-95, 15--19 May, 1995, Ecole Polytechnique, Palaiseau, France.}
}
\\
\vspace{0.4cm}
{\it  Theoretical Physics Institute, University of Minnesota,
\\
Minneapolis, MN 55455, USA}
\end{center}
\vspace*{1.2ex}
{\centerline{\large{(On work done in collaboration with M.~Shifman)}}}
\vspace*{1.2cm}

\begin{abstract}
I present a way of reconciling gauge coupling unification in minimal
supersymmetry with small $\alphasmz\approx0.11$ and discuss the
ensuing consequences.
\end{abstract}
\end{titlepage}
\setcounter{footnote}{0}
\setcounter{page}{2}
\setcounter{section}{0}
\newpage



\vspace{0.25in}
\noindent
In this talk~\cite{ms} I would like to make two points:
\begin{itemize}
\item
The value of $\alphasmz$ resulting from
exact gauge coupling
unification in the most popular version of the MSSM predicts the range
of $\alphasmz$ significantly above the range of rather low values,
not much above $0.11$, implied by
most experimental determinations and arguments from QCD; and
\item
One possible way of removing this discrepancy is to dramatically
alter the usual relation between the masses of the wino and the
gluino. This leads to important consequences for both phenomenology
and GUT-scale physics. One solid prediction of this approach is the
existence of a gluino below about 300\gev, within the reach of the
upgraded Tevatron.
\end{itemize}

\vspace{0.075in}
\noindent{{\underline{\em Calculating $\alphasmz$ in the MSSM.}}\hspace{0.05in}
It is well-known that, assuming that the gauge couplings unify,
one can predict one of them treating the other two as input.
Currently, most studies use the 2-loop renormalization group
equations (RGE's)
$
{d\alpha_i}/{d t} = {b_i}/(2\pi)\;\alpha_i^2
+ \mbox{two-loops},
$
where $i=1,2,3$,
$t\equiv\log(Q/\mz)$ and $\alpha_1\equiv\frac{5}{3}\alpha_Y$.
The one-loop
coefficients $b_i$ of the $\beta$ functions for the gauge couplings
change
across each new running mass threshold. Their parameterization in the
MSSM in the leading-log
approximation can be found, \eg, in~\cite{ekntwo,rr}.

The predicted value for $\alphasmz$ depends on the values of the
input parameters: $\alpha$, $\ssqthw(\mz)$, and $\mt$. It also
receives
corrections from: mass thresholds at the
electroweak scale, the  GUT-scale mass thresholds and
non-renormalizable operators,
the two-loop gauge and Yukawa contributions, and from scheme
dependence ($\MS$ {\it vs.} $\DR$).

For the electromagnetic coupling we
take~\cite{pdb} $\alpha^{-1}(\mz)=127.9\pm0.1$.
Recently, three groups have reanalyzed
$\alpha(\mz)$~\cite{alpha_recent}
and obtained basically similar results which do not change the resulting
value of $\alphasmz$ significantly~\cite{lp:new,ms}.

The input value of $\ssqthw(\mz)$ is
critical. This sensitivity is due to the fact that  $\alpha_2(Q)$ does
not change  between
$Q=\mz$ and the GUT scale $Q=\mgut$ as much as the other two
couplings.
Thus, a small increase in $\ssqthw(\mz)$ has an enhanced (and
negative)
effect on
the resulting value of $\alphasmz$.
Following Ref.~\cite{lp:new}
we assume
\beq
\ssqthw(\mz)=0.2316\pm0.0003 - 0.88\times10^{-7}{\gev}^2
\left[\mt^{2}
- (160\gev)^{2} \right].
\label{s2winput:eq}
\eeq
The  global analysis
of  Ref.~\cite{EL} implies that in the MSSM
$\mt=160\pm13\gev$, consistent with the recently reported~\cite{expt:top}
discoveries of the top quark:
$\mt=176 \pm8\pm10\gev$ (CDF) and $\mt= 199\pm
20\pm22\gev$. Taking, instead of 160\gev, even the D0 central value
for $\mt$
would lower $\ssqthw(\mz)$ and {\em increase}
$\alphasmz$ by only 0.005. Shifting $\ssqthw(\mz)$ up/down by 0.0003
shifts $\alphasmz$ down/up by only 0.0013.

Two-loop contributions in the RGE's increase $\alphasmz$ by
about 10\%. The most important among them is the pure
gauge term which yields $\Delta\alphasmz=0.012$ if one
assumes
SUSY in both one-  and two-loop  coefficients of the $\beta$ function
all the way down
to $Q=\mz$. Other corrections, from two-loop thresholds, Yukawa
contributions, and due to changing from the conventional $\MS$
scheme used here to the fully supersymmetric
$\DR$ scheme, are much smaller~\cite{lp:new,ms}.

\vspace{0.075in}
\noindent{{\underline{\em $\alphasmz$ from Constrained MSSM.}}\hspace{0.05in}
The MSSM, treated merely as the supersymmetrized Standard Model,
contains a multitude of free parameters. One
expects it to be a low-energy effective theory resulting from
some GUT or, more generally, some more fundamental scenario (e.g. strings)
valid at scales very much larger then the electroweak scale.
Depending on one's preferences for a more fundamental theory, one
can make various additional assumptions relating
the free parameters of the MSSM.

Perhaps the most commonly made assumption, other than the assumption
of gauge unification itself, is the relation among gaugino masses:
$M_1(\mgut)=M_2(\mgut)=\mgluino(\mgut)\equiv\mhalf$ which assigns at the
GUT scale the same
[common gaugino] mass to the bino, wino, and gluino states,
respectively.
This leads, due to renormalization
effects, to the following well-known relations at the electroweak scale:
\bea
\mone&=& {5\over3}\tan^2\theta_{\rm W}\,\mtwo\simeq0.5 \mtwo,
\label{monemtwo:eq} \\
\mtwo &=& \frac{\alpha_2}{\alphas}\mgluino\simeq 0.3\,\mgluino.
\label{mtwomgluino:eq}
\eea
These relations, or at least the first of them,
are commonly assumed in most
studies of the MSSM, even though, strictly speaking, they are not
necessary in the context of the model.

Another
relation which stems from minimal supergravity
and which is commonly
assumed is the equality of all the (soft) mass parameters of all the sleptons,
squarks, and typically also
Higgs bosons, to some [common scalar] mass parameter $\mzero$
at the GUT scale.
Renormalization effects
cause the masses of
color-carrying sparticles to become, at the $\mz$ scale,
typically by a factor of a few
heavier than the ones of the states with electroweak interactions only.

Often one also imposes a very attractive mechanism
of radiative electroweak symmetry breaking (EWSB), which provides additional
constraint on the parameters of the model, in particular relates the
SUSY Higgs/higgsino mass parameter $\mu$ to the parameters
of the model which break SUSY. This fully constrained framework is sometimes
called the Constrained MSSM (CMSSM)~\cite{kkrw1}. In practice,
various groups have considered the MSSM
with a varying number of additional assumptions, starting from
adopting just the common gaugino mass
to the CMSSM with additional constraints,
\eg, from nucleon decay which requires specifying the underlying GUT,
or string, model, the simplest $SU(5)$ and $SO(10)$
models being the most commonly studied. (A discussion of GUT physics
is beyond the scope of
this talk and I will only occasionally make references to expected
corrections to low-energy variables, like $\alphasmz$, from simplest
GUT-models.) It is not always easy to
discern what assumptions are actually responsible for what results.

At the level of accuracy described above, several
studies~\cite{rr,roberts,kkrw1,bbo,lp1,lp:new}
have generally agreed that, in the MSSM with additional assumptions of
the common gaugino and scalar masses, and
if one restricts oneself to masses roughly
below 1\tev\ then
$\alphasmz\gsim0.12$ (and $\gsim0.13$ for SUSY
masses below some 300\gev).
This is because $\alphasmz$ grows
with decreasing masses of SUSY particles.
The above prediction for  $\alphasmz$ has been considered  a
success and the strongest evidence
in favor of supersymmetric unification,
especially in light of the range of $\alphasmz=0.127\pm0.05$
resulting from the Z line shape at LEP~\cite{EL}.\\

\vspace{0.075in}
\noindent{{\underline{\em Is There an $\alphas$
Problem?}}\hspace{0.05in}
In spite of the existence of its several
independent determinations, there is still no
consensus on the experimental value of $\alphasmz$.
Most low-energy measurements and
lattice calculations of $\alphas$, when
translated to the scale $\mz$, generally give values much
below the LEP range, between
0.11 and 0.117, with comparable or smaller error bars. (See, \eg,
recent reviews~\cite{langacker} for more detail.)
The only indication (with small error bars) from low energies
for larger $\alphasmz$ from $\tau$ decays~\cite{pich} has been
questioned~\cite{shifman}.  A general
tendency among various reviews on the value of $\alphas$ is to
acknowledge the apparent discrepancy but adopt a sort of
``wait-and-see'' approach. Indeed, the world-average of
$0.117\pm0.006$ (see Bethke in Ref.~\cite{langacker})
is at least marginally consistent with most
determinations, even though they seem to correspond to two
disconnected (at $1\sigma$) sets of values. In fact, three
speakers~\footnote{One could say: As many as three speakers!}
at this meeting have assured us that, in this sense, there is no real
$\alphas$ problem.

However, recently Shifman~\cite{shifman}
very vigorously argued that
the internal consistency of QCD requires that $\alphas$ be close to 0.11.
He gave some important reasons. Here I will quote only one:
large $\alphas$ $\approx0.125$ would correspond to
$\Lambda_{\MS}\approx500\mev$ (in contrast to $\sim200\mev$ for
$\alphasmz\approx0.11$). Such a large value is apparently in conflict with
crucial features of QCD on which a variety of phenomena depend
sensitively.
Prompted by Shifman's argument,
Voloshin~\cite{voloshin} re-analyzed
$\Upsilon$ sum rules
claiming the record accuracy achieved so far: $\alphasmz = 0.109\pm
0.001$. Also,
a recent global fit~\cite{ksw} to LEP data favors $\alphasmz =0.112$.
Clearly, small $\alphasmz\approx0.11$ seems an increasingly viable
possibility, while significantly larger values are predicted by the
CMSSM.\\

\vspace{0.075in}
\noindent{{\underline{\em Can SUSY Unification Be Made Consistent with Small
$\alphasmz\approx0.11$?}}\hspace{0.05in}
Clearly, in the absence of large corrections from GUT-physics,
the most popular version of the MSSM, with the additional mass
relations between the gauginos and the scalars predicts too large values
of $\alphasmz$.

Several solutions to this problem can be immediately suggested.
One is to remain
in the context of the CMSSM but adopt
a heavy SUSY scenario with the SUSY mass spectra significantly
exceeding
1\tev. This scenario would not only put SUSY into both theoretical
and experimental oblivion,
but is also, for the most part, inconsistent
with
our expectations that the lightest supersymmetric particle (LSP)
should be neutral
and/or with the lower bound on the age
of the Universe of at least some 10 billion
years. (See Section~5 of Ref.~\cite{kkrw1}.)
Another possibility is to invoke large enough
negative corrections due to GUT-scale physics.
The issue was re-analyzed recently~\cite{lp:new} and it was found that,
under natural assumptions about the scale of GUT-scale corrections,
$\alphasmz>0.12$.
Although it may well happen that the GUT-scale corrections
are abnormally large, the predictive power
is essentially lost. One can also employ~\cite{mohapatra}
an intermediate scale for
which there is good motivation from neutrino and axion physics.

Here, I would like to present a different approach. Its starting point is
the question: Which ingredient of the CMSSM is mainly responsible for
predicting large $\alphasmz>0.12$? Is it the
particle content of the MSSM, or rather some additional assumptions, which
may not be a necessary ingredient of the model, and could therefore be
relaxed.

To answer this question, in the first step let's treat all the mass
parameters entering [threshold corrections in] the RGE's as completely
unrelated from each other. Let's thus assume no relations of any kind
between the gauginos or between the scalars. (Since GUT-scale corrections are
GUT model-dependent, let's also turn them off, while keeping in mind that
in reality they may be sizable, and that they may both contribute to
increasing and decreasing $\alphasmz$.) By requiring only that all the masses
be less than about 1\tev, for comparison with the CMSSM case, we find that

\beq
\alphasmz>0.106
\label{alphasmzmin:eq}
\eeq

\noindent
This shows that, in the MSSM itself, without imposing
any additional mass relations,
one can in principle easily obtain much
smaller $\alphasmz$ than in the CMSSM. Since our task is
to minimize $\alphasmz$, we take $\mt=160\gev$ while fixing
$\ssqthw(\mz)$ at its central value.
Of course, the range of $\alphasmz$ values resulting from [exact]
supersymmetric unification, still depends on the MSSM mass parameters.
We thus arbitrarily
choose them in such a way as
to further minimize $\alphasmz$: we set ${M_2}$, ${m_{\tilde{l}}}$,
${m_{{\tilde t}_L}}$, $m_{\widetilde H}$, and $m_{H_2}$ at
1\tev, and
${m_{\widetilde g}}$, ${m_{\tilde{q}}}$, and ${m_{{\tilde t}_R}}$ at 100\gev.
(See~\cite{ms} for more details.)

How can such small values of $\alphasmz$ be consistent with supersymmetric
unification?
It can be easily seen that it is the gluino and the wino that play the
dominant role in influencing $\alphasmz$.
This can be done by examining the form of the
1-loop mass  threshold coefficients $b_i$ entering the RGE's.
Not only are the mass threshold corrections due to both the wino and the
gluino the largest ($4/3$ and 2, respectively)
but also each of them affects the running of only one
coupling, $\alpha_2$ and $\alphas$, respectively.
By lowering $\mgluino$ and increasing $\mtwo$ one can easily descend
to small $\alphasmz$ in the vicinity of 0.11. Shifting other masses
has a much smaller impact on the resulting range of $\alphasmz$. (See
Fig.~1 of~\cite{ms}.)

\begin{figure}
\centering
\epsfxsize=3in
\hspace*{0in}
\epsffile{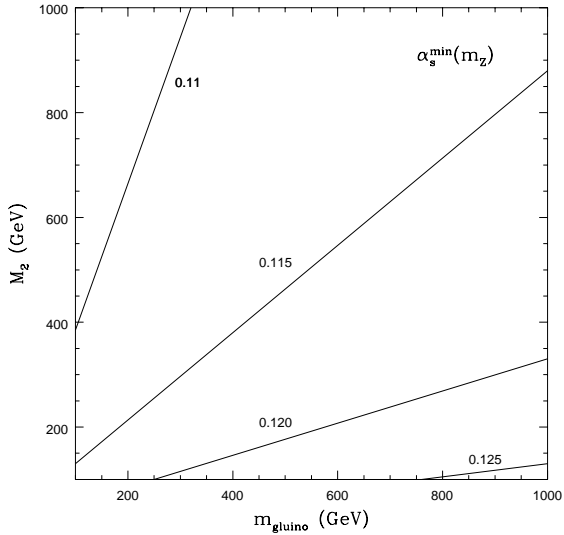}
\caption{Contours of constant $\alphasmzmin$ in the
($\mgluino,\mtwo$)
plane. All other mass parameters are chosen so as to minimize
$\alphasmz$ (see text) and  $\mt=160\gev$.
}
\label{winogluino:fig}
\end{figure}

\vspace{0.075in}
\noindent{{\underline{\em Implications.}}\hspace{0.05in}
The price that one has to pay is evident and can be seen in
Fig.~\ref{winogluino:fig}. Without invoking other
effects, like large GUT-scale corrections, one has to have the wino
actually heavier than the gluino. If $\alphasmz\approx0.11$ and
no large negative GUT-scale corrections are invoked, then
$\mgluino\lsim300\gev$ and $\mtwo\gsim400\gev$. In fact one finds
$\mtwo\gsim3\mgluino$, thus violating the
relation~(\ref{mtwomgluino:eq}). (In order for the lightest neutralino
to weight less than the gluino, and thus be the LSP, also the
relation~(\ref{monemtwo:eq}) has to be violated.)
This is a clear
prediction which, remarkably, can be tested even before the end of the
millennium. If the upgraded Tevatron, which
will search for gluinos up to about $300\gev$, finds a gluino below
some $240\gev$, while LEP~II does not find a wino-like chargino up to
about $80\gev$, then~(\ref{mtwomgluino:eq}) will likely be violated. (The
chargino could still be of higgsino type but then so would be the lightest
neutralino which would leave the MSSM without
a viable candidate for DM in the
Universe~\cite{dmafterleptwo}.)
A vast majority of studies of all aspects of supersymmetry routinely
assume~(\ref{monemtwo:eq})--(\ref{mtwomgluino:eq}).

There are also other important implications of this surprising
scenario which have to do with some long-lasting anomalies in the
$b\bar b$ system~\cite{ms}.

Finally, what implications for physics structure at the GUT scale follow from
this bottom-up approach? In standard GUT's equal gaugino
masses at $\mgut$ are enforced by GUT gauge invariance. This is the
most popular scenario in which equal gaugino masses are generated by
coupling a SUSY GUT to $N=1$ minimal supergravity (SUGRA) and choosing the
simplest, delta-function, kinetic term for the gauge/gaugino fields.
If instead one considers a general form of the kinetic term, one finds
that relations among gaugino masses become arbitrary which could
potentially give room to accommodate large gaugino mass non-degeneracy.
However, it comes out that in this general case of
non-minimal SUGRA also the gauge
couplings become unequal at $\mgut$~\cite{eent}. Since in our analysis
we assume them equal (for the reasons of
simplicity), we could attempt to implement the resulting gaugino mass
ratios
in non-minimal SUGRA by allowing some small blurring in gauge coupling
unification (due to some small GUT-scale corrections) and still trying to
generate large split in gaugino masses needed for lowering $\alphasmz$.
However, at least in simplest GUT models (like $SU(5)$ and $SO(10)$)
coupled to $N=1$ SUGRA, it is probably
impossible to accommodate such a large non-universality of gaugino
masses while preserving almost exact gauge coupling
unification~\cite{dasgupta}.
In this type of scenarios (even with general
kinetic terms) the mechanism of lowering $\alpha_s(m_Z)$ presented
here may play at best only a subdominant role. On the other hand, it
may better fit an alternative approach in which
non-zero gaugino masses are only generated below
$\mgut$~\cite{dineetal}.

\vspace{0.075in}
\noindent{{\underline{\em Conclusions.}}\hspace{0.05in}
There are a host of indications that the true value of $\alphasmz$ may be
close to 0.11. The most popular form of the MSSM, with standard
gaugino mass relations, predicts $\alphasmz$ {\em at least} some 10\%
larger. In that scenario, one would have to invoke large and negative
GUT-scale corrections to rescue gauge coupling unification. This would
provide a stringent constraint on GUT models while, at the electroweak
scale, the predictive power would be significantly reduced.

We have
made a simple observation that, even in the MSSM, one is able to obtain
$\alphasmz$ in the vicinity of 0.11. This can be done
at the expense of abandoning the
usual mass relations among gaugino
masses~(\ref{monemtwo:eq})--(\ref{mtwomgluino:eq}).
One firm prediction of this
approach is the existence of a relatively light gluino,
$\mgluino\lsim300\gev$ and a heavy wino-like chargino,
$\mtwo\gsim3\mgluino$, which can be tested in the upgraded Tevatron
and LEP~II. A relatively light gluino, in the ballpark of
100\gev, may also help
solving some long-lasting puzzles in $b$-quark physics. Implications
for GUT-scale physics are also dramatic in the sense that this
solution implies a large non-degeneracy of gaugino masses at $\mgut$
in contrast with a standard approach in which GUT models are coupled
to $N=1$ supergravity.


\begin{thebibliography}{99}
\bibitem{ms}
L.~Roszkowski and M.~Shifman,
hep-ph/9503358, to appear in Phys. Rev. {\bf D}.

\bibitem{ekntwo}
J.~Ellis, S.~Kelley, and D.V.~Nanopoulos,
\PLB{260}{91}{131}.

\bibitem{rr}
R.G.~Roberts and G.G.~Ross, Nucl. Phys. {\bf B377} (1992) 571.

\bibitem{pdb}
L.~Montanet, {\em et al.} (PDG), Phys. Rev. {\bf D50} (1994) 1173.

\bibitem{alpha_recent}
A.~Martin and D.~Zeppenfeld,
Phys. Lett. {\bf B345} (1995) 558;
M.L.~Swartz,
hep-ph/9411353;
S.~Eidelman and F.~Jegerlehner,
hep-ph/9502298.

\bibitem{lp:new}
P.~Langacker and N.~Polonsky,
hep-ph/9503214.

\bibitem{EL}
J.~Erler and P.~Langacker,
hep-ph/9411203.

\bibitem{expt:top}
F.~Abe, {\em et al.} (CDF), 
hep-ex/9503002;	
S.~Abachi, {\em et al.} (D0), 
hep-ex/9503003.

\bibitem{roberts}
R.G.~Roberts and L.~Roszkowski, Phys. Lett. {\bf B309} (1993) 329.

\bibitem{kkrw1}
G.~Kane, C.~Kolda, L.~Roszkowski, and J.~Wells,
\PRD{49}{94}{6173}.

\bibitem{bbo}
V.~Barger, M.S.~Berger, and P.~Ohmann, \PRD{47}{93}{1093}.

\bibitem{lp1}
P.~Langacker and N.~Polonsky, \PRD{47}{93}{4028}.

\bibitem{langacker}
P.~Langacker,
hep-ph/9412361 and
hep-ph/9411247; S.~Bethke, at Rencontres de Moriond, les Arcs, March 1995.

\bibitem{pich}
See, \eg, F.~Le~Diberder and A.~Pich, \PLB{286}{92}{147}.

\bibitem{shifman}
M.~Shifman,
Mod. Phys. Lett. {\bf A10} (1995) 605.

\bibitem{voloshin}
M.~Voloshin,
\IJMPA{10}{95}{2865}.

\bibitem{ksw}
G.~Kane, R.~Stuart, and J.~Wells, hep-ph/9505207; G.~Kane,
talk at this meeting.

\bibitem{mohapatra}
D.-G.~Lee and R.N.~Mohapatra,
hep-ph/9502210;
B.~Brahmachari and R.N.~Mohapatra, 
hep-ph/9505347 and 
hep-ph/9508293.

\bibitem{dmafterleptwo}
L.~Roszkowski, {\rm Phys. Lett.} {\bf B262},
59 (1991) and {\rm Phys. Lett.} {\bf B278},
147 (1992).

\bibitem{eent}
J.~Ellis, K.~Enqvist, D.~Nanopoulos, and K.~Tamvakis,
Phys. Lett. {\bf 155B} (1985) 381.

\bibitem{dasgupta}
T.~Dasgupta, P.~Mamales, and P.~Nath,
hep-ph/9501325;
R.~Arnowitt, private communication; K.~Kounnas, private communication.

\bibitem{dineetal}
M.~Dine, A.E.~Nelson, Y.~Nir, and Y.~Shirman,
hep-ph/9507378.

\end{thebibliography}
\end{document}